\documentclass[12pt]{article}

\usepackage{latexsym}

\usepackage{graphicx}

\begin{document}


\title{Rotating non-asymptotically flat black rings in charged dilaton gravity }

\author{
     Stoytcho S. Yazadjiev \thanks{E-mail: yazad@phys.uni-sofia.bg}\\
{\footnotesize  Department of Theoretical Physics,
                Faculty of Physics, Sofia University,}\\
{\footnotesize  5 James Bourchier Boulevard, Sofia~1164, Bulgaria }\\
{\footnotesize and }\\
{\footnotesize  Bogoliubov Laboratory of Theoretical Physics,
                JINR, 141980 Dubna, Russia}\\
}

\date{}

\maketitle

\begin{abstract}
We derive new rotating, non-asymptotically flat black ring solutions in five-dimensional
Einstein-Maxwell-dilaton gravity with dilaton coupling constant $\alpha=\sqrt{8/3}$ which
arises from a six-dimensional Kaluza-Klein theory. As a limiting case we also find  new
rotating, non-asymptotically flat five-dimensional  black holes. The solutions are analyzed
and the mass, angular momentum and charge are computed. A Smarr-like relation is found.
It is shown that the first law of black hole thermodynamics is satisfied.
\end{abstract}


\sloppy

\section{Introduction}

An interesting development in the black holes studies is the discovery of the black ring solutions
of the five-dimensional Einstein equations by Emparan and Reall \cite{ER1}, \cite{ER2}. These are
asymptotically flat solutions with an event horizon of topology $S^2\times S^1$ rather the
much more familiar $S^3$ topology. Moreover, it was shown in \cite{ER2} that both the black hole
and the the black ring can carry the same conserved charges, the mass and a single angular
momentum, and therefore there is no uniqueness theorem in five dimensions. Since the Emparan and
Reall's discovery many explicit examples of black ring solutions were found in various gravity
theories \cite{E}-\cite{P}. Elvang was able
to apply Hassan-Sen transformation to the solution \cite{ER2} to find a charged black ring in the
bosonic sector of the truncated heterotic string theory\cite{E}. A supersymmetric black ring in
five-dimensional minimal supergravity was derived in \cite{EEMR1} and then generalized to the
case of concentric rings in \cite{GG1} and \cite{GG2}. A static black ring solution of the five
dimensional Einstein-Maxwell gravity was found by Ida and Uchida in \cite{IU}. In \cite{EMP}
Emparan derived "dipole black rings" in Einstein-Maxwell-dilaton (EMd) theory
in five dimensions. In this work Emparan showed that the black rings can exhibit novel
feature with respect to the black holes. The black rings can also carry nonconserved charges
which can be varied continuously without altering the conserved charges. This fact leads to
continuous non-uniqness. The thermodynamics of the dipole black rings, within the quasilocal
counterterm method, was discussed by Astefanesei and Radu \cite{AR}.
Static and asymptotically flat black rings solutions in five-dimensional
EMd gravity with arbitrary dilaton coupling parameter $\alpha$ were presented in \cite{KL}.
Non-asymptotically flat black rings immersed in external electromagnetic fields were found
and discussed  in \cite{O} and \cite{KL}.A systematical derivation of the asymptotically
flat static black ring solutions
in five-dimensional EMd gravity with an arbitrary dilaton coupling parameter was given in
\cite{Y}. In the same paper, the author systematically derived new type static black ring
solutions which are not asymptotically flat.

The aim of this paper is to generalize the results
of \cite{Y} and to present exact rotating  non-asymptotically flat black rings solutions
in five-dimensional EMd gravity. Such type EMd solutions, however in the case of black holes
(i.e. with spherical horizon topology), have attracted much interest in recent years
\cite{PW}-\cite{Y1}.

In general, the $D$-dimensional EMd gravity is described by
the action\footnote{In what follows we consider  theories with $\alpha\ne 0$.}

\begin{equation}\label{EMDA}
S= {1\over 16\pi} \int d^Dx \sqrt{-g}\left(R - 2g^{\mu\nu}\partial_{\mu}\varphi \partial_{\nu}\varphi  -
e^{2\alpha\varphi}F^{\mu\nu}F_{\mu\nu} \right).
\end{equation}

The field equations derived from the action are

\begin{eqnarray}
R_{\mu\nu} &=& 2\partial_{\mu}\varphi \partial_{\nu}\varphi + 2e^{2\alpha\varphi} \left[F_{\mu\rho}F_{\nu}^{{\,}\rho} - {g_{\mu\nu}\over 2(D-2)} F_{\beta\rho} F^{\beta\rho}\right], \\
\nabla_{\mu}\nabla^{\mu}\varphi &=& {\alpha\over 2} e^{2\alpha\varphi} F_{\nu\rho}F^{\nu\rho}, \\
&\nabla_{\mu}&\left[e^{2\alpha\varphi} F^{\mu\nu} \right]  = 0 .
\end{eqnarray}

\section{Static, non-asymptotically flat black rings}

In this section, following \cite{Y} we present the derivation of the static
non-asymptotically flat black ring solutions.

The metric of the static spacetime (i.e. admitting the hypersurface orthogonal Killing
vector $\xi={\partial\over \partial t}$ ) can be written in the form

\begin{equation}
ds^2 = - e^{2U}dt^2 + e^{-{2U\over D-3 }} h_{ij}dx^idx^j
\end{equation}

where $U$ and $h_{ij}$ are independent of the time coordinate $t$.
The electromagnetic field is taken in the form

\begin{equation}
F = e^{-2U}\xi \wedge d\Phi
\end{equation}
where $\Phi$ is the electrostatic potential.

In terms of the potentials
$U$, $\Phi$, $\varphi$ and $(D-1)$-dimensional metric $h_{ij}$ the field equations become

\begin{eqnarray}
{\cal D}_{i}{\cal D}^{i}U &=& 2{D-3\over D-2 }e^{-2U+2\alpha\varphi}h^{ij}{\cal D}_{i}\Phi {\cal D}_{j}\Phi , \\
{\cal D}_{i}{\cal D}^{i}\varphi &=& -\alpha e^{-2U+2\alpha\varphi}h^{ij}{\cal D}_{i}\Phi {\cal D}_{j}\Phi ,\\
&{\cal D}_{i}&\left( e^{-2U+2\alpha\varphi}h^{ij}{\cal D}_{j}\Phi \right) = 0 ,\\
{\cal R}(h)_{ij} &=& {D-2\over D-3} {\cal D}_{i}U {\cal D}_{j}U + 2{\cal D}_{i}\varphi {\cal D}_{j}\varphi
- 2e^{2\alpha\varphi -2U}{\cal D}_{i}\Phi {\cal D}_{j}\Phi ,
\end{eqnarray}

where ${\cal D}_{i}$ and ${\cal R}(h)_{ij}$ are the covariant derivative and the Ricci tensor with respect to the metric $h_{ij}$. These equations can be derived from the action

\begin{equation}
S = \int d^{(D-1)}x \sqrt{h} \left[{\cal R}(h) - {D-2\over D-3}h^{ij}{\cal D}_{i}U {\cal D}_{j}U -
2h^{ij}{\cal D}_{i}\varphi {\cal D}_{j}\varphi    + 2e^{2\alpha\varphi - 2U}h^{ij}{\cal D}_{i}\Phi {\cal D}_{j}\Phi\right].
\end{equation}

Let us introduce the symmetric matrix

\begin{eqnarray}
P = e^{(\alpha_{D} -1)U} e^{ (\alpha_{D} + 1)\varphi_{D}}\left(%
\begin{array}{cc}
  e^{2U - 2\alpha_{D}\varphi_{D}} - (1 + \alpha^2_{D})\Phi^2_{D} & - \sqrt{1 + \alpha^2_{D}}\Phi_{D} \\
- \sqrt{1 + \alpha^2_{D}}\Phi_{D}  &  -1 \\\end{array}%
\right)
\end{eqnarray}

where

\begin{equation}
\alpha_{D}= \sqrt{{D-2\over 2(D-3)}} \alpha, \,\,\, \varphi_{D} = \sqrt{{2(D-3)\over (D-2)}} \varphi, \,\,\, \Phi_{D}= \sqrt{{2(D-3)\over (D-2)}} \Phi .
\end{equation}

Then the action can be written in the form

\begin{equation}
S = \int d^{(D-1)}x \sqrt{h} \left[{\cal R}(h) + {1\over 2(1+ \alpha_{D}^2) }{(D-2)\over (D-3)} h^{ij} Sp \left({\cal D}_{i}P {\cal D}_{i}P^{-1} \right)  \right].
\end{equation}

The action is invariant under the symmetry transformations

\begin{equation}
P \longrightarrow GPG^{T}
\end{equation}

where $G\in GL(2,R)$. The matrix $P$ parameterizes the coset $GL(2,R)/SO(1,1)$.
Similar sigma model was found in \cite{GR}. However, here we give the parameterization
of the target space in terms of $2\times 2$ matrices rather than $3\times 3$ matrices
as it is done in \cite{GR}.

Let us consider a static asymptotically flat  (vacuum) solution of $D$-dimensional
Einstein equations

\begin{equation}\label{SFDS}
ds_{0}^2 = -e^{2U_{0}}dt^2 + e^{-{2U_{0}\over D-3}}h_{ij}dx^idx^j
\end{equation}

which is encoded into the matrix
\begin{equation} P_{0}= e^{(\alpha_{D} -1)U_{0} }
\left(%
\begin{array}{cc}
  e^{2U_{0}} & 0 \\
0 & -1 \\\end{array}%
\right) .
\end{equation}

and the metric $h_{ij}$. In order to generate non-asymptotically flat
EMd solutions we shall consider the matrix

\begin{equation}
N = \left(%
\begin{array}{cc}
 0 & - a^{-1} \\
 a & a \\\end{array}%
\right) \in SL(2,R)\subset GL(2,R) .
\end{equation}

Then we obtain the following EMd solution presented by the matrix

\begin{equation}\label{NAFT}
P = N P_{0} N^{T}
\end{equation}

i.e.

\begin{eqnarray}\label{NAFEMDS}
e^{U} &=& {e^{U_{0}}\over \left[a^2(1 - e^{2U_{0}}) \right]^{1\over 1 + \alpha^2_{D} } }
\nonumber, \\
e^{\varphi_{D}} &=& \left[a^2(1 - e^{2U_{0}}) \right]^{\alpha_{D}\over 1 + \alpha^2_{D} } ,\\
\Phi_{D} &=& - {a^{-2} \over \sqrt{1 + \alpha^2_{D} } } {1\over 1 - e^{2U_{0}} } \nonumber.
\end{eqnarray}

As an explicit example of a seed vacuum solution we consider the five-dimensional,
static, asymptotically flat black ring solution given by the metric

\begin{eqnarray}\label{BRS}
ds^2_{0} &=& - {F(y)\over F(x)}dt^2 \\ &+& {{\cal R}^2\over (x-y)^2}
\left[F(x)(y^2-1)d\psi^2 + {F(x)\over F(y)} {dy^2\over (y^2-1)}
 + {dx^2\over (1-x^2) } + F(x) (1-x^2) d\phi^2\right] \nonumber
\end{eqnarray}

where\footnote{We warn the reader that even if we use the same letters as in \cite{Y} their
meaning is different. Here we present the black ring solutions in the coordinates of \cite{EMP}.}
$F(x)= 1 + \lambda x$, ${\cal R}>0$ and $0<\lambda < 1$.
The coordinate $x$ is in the range $-1\le x\le 1$ and the
coordinate $y$ is in the range $-\infty<y\le -1$.
The solution has a horizon at $y=-1/\lambda$ . The topology of the horizon is $S^2\times S^1$,
parameterized by $(x,\phi)$ and $\psi$, respectively. In order to avoid a conical singularity
at $y=-1$ one must demand that the period of $\psi$
satisfies $\Delta \psi=2\pi/\sqrt{1 - \lambda}$. If one demands regularity at $x=-1$ the period of
$\phi$ must be $\Delta \phi=2\pi/ \sqrt{1 -\lambda}$. In this case the solution is
asymptotically flat and the ring is sitting on the rim of disk shaped membrane with
a negative deficit angle.
To enforce regularity at $x=1$ one must take $\Delta \phi = 2\pi/ \sqrt{1+\lambda}$
and the solution describes a black ring sitting on the rim of disk shaped hole
in an infinitely extended deficit membrane with positive deficit.
More detailed analysis of the black ring solution can be found in \cite{ER1}.

Applying the transformation (\ref{NAFT})to the neutral black ring solution (\ref{BRS})
we obtain the following EMd solution

\begin{eqnarray}\label{SNFEMDS}
ds^2 &=& - \left[a^2\left(1 - {F(y)\over F(x) }\right) \right]^{{-2\over 1 + \alpha^2_{5}}}
 {F(y)\over F(x)}dt^2 \nonumber \\
&+& {{\cal R}^2\left[a^2\left(1 - {F(y)\over F(x) }\right) \right]^{{1\over 1 + \alpha^2_{5}}} \over (x-y)^2 }
\left[ F(x)(y^2-1)d\psi^2 + {F(x)\over F(y)}{dy^2\over y^2 -1}
 + {dx^2\over 1-x^2 } + F(x)(1-x^2)d\phi^2\right] \nonumber \\
e^{\varphi_{5}} &=& \left[a^2\left(1 - {F(y)\over F(x) }\right) \right]^{\alpha_{5}\over 1 + \alpha^2_{5} } ,\\
\Phi_{5} &=& - {a^{-2} \over \sqrt{1 + \alpha^2_{5} } } {1\over 1 - {F(y)\over F(x)} } \nonumber .
\end{eqnarray}

From the explicit form of the solution it is clear that there is a horizon at $y=-1/\lambda$ with
topology $S^2\times S^1$. Using arguments similar to that in \cite{ER1} (see also \cite{Y1}) one can show
that the horizon is regular and the solution can be continuously extended through the horizon.

The analysis of the near-horizon geometry is quite similar to that for the near-horizon
geometry of the neutral black ring solution. As it is seen the both near-horizon
geometries are conformal with a constant conformal factor. In order for the metric to
be regular at $x=-1$ the period of $\phi$ must be $\Delta \phi=2\pi/ \sqrt{1 - \lambda}$.
The regularity at $x=1$ requires $\Delta \phi=2\pi/ \sqrt{1+\lambda}$. Therefore, it is not
possible for the metric to be regular at both $x=-1$ and $x=1$.
The regularity at $x=-1$ means that there is a conical singularity
at $x=1$ and vice versa. The $x\phi$ part of the metric describes a
two-dimensional surface with $S^2$ topology  and with a conical singularity at one of the poles.

The asymptotic infinity corresponds to $x=y=-1$. One can show that near $x=y=-1$ the
solution behaves as

\begin{eqnarray} \label{SBRAM}
ds^2 &\approx& - \left[a^{-2}{(1-\lambda)\over 2\lambda } {r^2\over {\cal R}^2}\right]^{2\over 1+\alpha^2_{5} } dt^2
\\
&+& \left[a^{-2}{(1-\lambda)\over 2\lambda }
 {r^2\over {\cal R}^2}\right]^{-1\over 1+\alpha^2_{5} } \left[dr^2 +
 r^2d\theta^2 + r^2\cos^2\theta d{\tilde \psi}^2  + r^2\sin^2\theta d{\tilde \phi}^2  \right] ,\nonumber
\end{eqnarray}

\begin{eqnarray}\label{SBRAD}
e^{2\alpha\varphi} &\approx &\left[a^{-2}{(1-\lambda)\over 2\lambda }
{r^2\over {\cal R}^2}\right] ^{2\alpha^2_{5}\over 1+\alpha^2_{5} } ,
\end{eqnarray}

\begin{eqnarray}\label{SBRAP}
\Phi &\approx&- {\sqrt{3}\over 2 \sqrt{1+ \alpha^2_{5}} }
\left[a^{-2}{(1-\lambda)\over 2\lambda } {r^2\over {\cal R}^2}\right] .
\end{eqnarray}

where the coordinates $r$, ${\tilde \psi}$ and ${\tilde \phi}$ are
defined as follows

\begin{equation}
r\cos\theta = {\cal R} {\sqrt{y^2 -1}\over x-y },\,\,\,
r\sin\theta = {\cal R} {\sqrt{1-x^2}\over x-y },\,\,\,
{\tilde \psi} = \sqrt{1-\lambda}\psi ,\,\,\, {\tilde \phi}= \sqrt{1- \lambda}\phi.
\end{equation}

It is clear that the solution is not asymptotically flat. Moreover, the presented asymptotic matches
the asymptotic of the spherically symmetric Chan-Horne-Mann solution of the five-dimensional EMd gravity
\cite{CHM}. It is worth noting that the $D$-dimensional Chan-Horne-Mann solution can be obtained via the solution generating transformations  (\ref{NAFEMDS}) applied to the $D$-dimensional Schwarzschild solution. In five dimensions, the black ring solution (\ref{BRS}) asymptotes ( $r\to \infty$) the Schwarzschild solution and therefore
the  transformations  (\ref{NAFEMDS}) generate the same asymptotic as for the Chan-Horne-Mann solution.

Even though our solution is not asymptotically flat and the dilaton field behaves like $\varphi \sim \ln(r)$
for large $r$,  the Ricci and the Kretschmann scalars vanish as $r\to \infty$. More precisely we have (see Appendix \ref{RKSSS})

\begin{equation}
R \sim  r^{-{2\alpha^2_{5} \over 1 + \alpha^2_{5}}}  ,\,\,\,\,
R_{\mu\nu\alpha\beta}R^{\mu\nu\alpha\beta}\sim r^{-{4\alpha^2_{5} \over 1 + \alpha^2_{5}} }.
\end{equation}

Therefore the spacetime is well-behaved for $r\to \infty$.

The horizon area is

\begin{equation}
{\cal A}_{h} = 8\pi^2 {\cal R}^3
{\lambda^2 a^{3\over (1 + \alpha^2_{5})}\over (1-\lambda)\sqrt{(1+\lambda)(1 \pm \lambda)} }.
\end{equation}

where the sign $\pm$ corresponds to taking the conical singularity at $x=\pm 1$.

The black ring mass, charge and temperature are given by

\begin{eqnarray}
M_{\pm} &=& {\alpha^2_{5}\over 1 + \alpha^2_{5}} {3\pi {\cal R}^2\over 4 }
{\lambda \over \sqrt{(1-\lambda)(1\pm \lambda)}  } ,\\
Q_{\pm} &=& - {\sqrt{3}\pi a^2{\cal R}^2\over \sqrt{1+\alpha^2_{5}}}
{\lambda \over \sqrt{(1-\lambda)(1\pm \lambda)}  } ,\\
T &=& {\sqrt{1-\lambda^2} \over 4\pi \lambda {\cal R} }a^{-{3\over (1 +\alpha^2_{5}) }}.
\end{eqnarray}

In order to compute the mass of the solution we have used the so-called quasilocal formalism.
Some details are given in  Appendix \ref{QLF}.

The electric potential is defined up to an arbitrary additive constant.
In the asymptotically flat case
there is a preferred gauge in which $\Phi(\infty)=0$. In the non-asymptotically
flat case the electric potential diverges at spacial infinity and there is no preferred gauge.
The arbitrary constant, however, can be fixed so that
the Smarr-type relation to be satisfied:

\begin{equation}
M_{\pm} = {3\over 8} TA_{h\pm} + \Phi_{h}Q_{\pm}.
\end{equation}

\section{Rotating, non-asymptotically flat black rings}

The five-dimensional EMd gravity with the dilaton coupling parameter $\alpha^2=8/3$
can be obtained as a dimensional reduction of the six dimensional vacuum Einstein gravity

\begin{equation}\label{SDEG}
S = \int d^6x \sqrt{-g_{(6)}}R_{(6)}
\end{equation}

along the spacelike Killing vector $\partial/\partial x^6 $. The Kaluza-Klein ansatz

\begin{equation}
ds^2_{6} = e^{- {2\varphi\over \sqrt{6}} } ds^2_{5}
+ e^{2\sqrt{{3\over 2}}\varphi}\left(dx^6 + 2A_{\mu}dx^\mu\right)^2
\end{equation}

reduces the action (\ref{SDEG}) to the five dimensional EMd gravity action (\ref{EMDA})
with $\alpha=\sqrt{8/3}$. The fact that the five-dimensional EMd theory with $\alpha=\sqrt{8/3}$
and the six-dimensional vacuum Einstein gravity are related via a Kaluza-Klein reduction is
important and enables us to generate exact solutions to EMd theory from solutions of the
six dimensional vacuum Einstein equations. Moreover, it is known that the stationary vacuum
Einstein equations in $D$ dimensions (i.e. with $D-3$ commuting Killing vectors) reduce to
$(D-2)$-dimensional self-gravitating $SL(D-2,R)/SO(D-2)$ sigma-model \cite{M},\cite{DM}.
Applying the $SL(D-2,R)$ symmetry transformations one can, in principle, generate new solutions
from known ones. This implies a natural and simple method for generating exact solutions to
the five-dimensional($(D-1)$-dimensional in general) EMd equations with $\alpha =\sqrt{8/3}$
(with $\alpha = \sqrt{2(D-2)/(D-3)}$ in general). The method consists in the following steps.
Given a $(D-1)$-dimensional vacuum Einstein solution is first trivially embedded
in a $D$-dimensional spacetime and then is $SL(D-2,R)$ transformed. After Kaluza-Klein reduction
one obtains a $(D-1)$-dimensional EMd solution. The described method was successfully applied to
many specific problems \cite{GW}-\cite{HH},\cite{CL}(and references therein).
In our situation the picture is the following. First, the $SL(2,R)$ transformation (\ref{NAFT})
with matrix $N$ is uplifted to $SL(4,R)$ transformation with matrix $S_{N}$. Then we apply
the so-obtained
$SL(4,R)$ transformation to the trivially embedded rotating neutral black ring solution. The new
spacetime is Kaluza-Klein reduced and the found EMd solution represents a rotating
non-asymptotically flat five-dimensional EMd black ring.
The technical details are presented below.

The metric of six-dimensional spacetime admitting three commuting Killing vectors

\begin{equation}
\xi = {\partial\over \partial t},\,\,\, \eta = {\partial\over \partial \psi}, \,\,\,
\chi = {\partial\over \partial x^6},
\end{equation}

can be written in the form

\begin{equation}
ds_{6}^2 = f_{ab}(dx^a + w^a_{i}dx^i)(dx^b + w^b_{j}dx^j) + f^{-1}H_{ij} dx^idx^j
\end{equation}

where $f=|\det(f_{ab})|$, $i=1,2,3$, $a=4,5,6$ ($x^4=t, x^5=\psi$). Let us introduce the twist
potentials
\begin{equation}
V_{a}={f f_{ab}\over \sqrt{H} }H_{ij}\varepsilon^{jmn}\partial_{n}w^b_{m}
\end{equation}

where $H=\det(H_{ij})$. The Einstein equations can then be written  as a three dimensional
selfgravitating sigma model

\begin{eqnarray}
R(H)_{ij} &=& {1\over 4} Sp\left[M^{-1}{\cal D}_{i}M M^{-1}{\cal D}_{j}M\right], \\
&{\cal D}_{i}&\left[M^{-1}{\cal D}^{i}M\right] = 0.
\end{eqnarray}

Here ${\cal D}_{i}$ is the coderivative operator and $R(H)_{ij}$ is the Ricci tensor
with respect to the metric $H_{ij}$. $M$ is a unimodular symmetric matrix given by

\begin{equation} M =
\left(%
\begin{array}{cc}
  f_{ab} - f^{-1}V_{a}V_{b} & - f^{-1}V_{a} \\
- f^{-1}V_{b} & - f^{-1}\\
\end{array}%
\right) .
\end{equation}

The field equations therefore are invariant under $SL(4,R)$ transformations

\begin{equation}\label{SLT}
M \longrightarrow SMS^T .
\end{equation}

Note that the three-metric $H_{ij}$ does not change under these transformations.

Let us consider  the trivial embedding of the five-dimensional static solution (\ref{SFDS})

\begin{equation}
ds_{6}^2 = -e^{2U_{0}}dt^2 + e^{-U_{0}}h_{ij}dx^idx^j + dx^2_{(6)}.
\end{equation}

The corresponding matrix $M$ is

\begin{eqnarray}
 M^0_{static} = \left(%
\begin{array}{cccc}
 -e^{2U_{0}} & 0 & 0 & 0 \\
  0 & 1 & 0 & 0 \\
  0 & 0 &  e^{-U_{0}}h_{\psi\psi} & 0 \\
  0 & 0 & 0 & -f^{-1}_{0}
\end{array} %
\right) .
\end{eqnarray}

The Kaluza-Klein embedding of (\ref{NAFEMDS}) has the representative matrix

\begin{eqnarray}
 M_{static} = \left(%
\begin{array}{cccc}
 {1\over a^2} & -1 & 0 & 0 \\
  -1 & a^2(1 -e^{2U_{0}}) & 0 & 0 \\
  0 & 0 &  e^{-U_{0}}h_{\psi\psi} & 0 \\
  0 & 0 & 0 & -f^{-1}_{0}
\end{array} %
\right) .
\end{eqnarray}

It can be checked that

\begin{equation}
M_{static} = S_{N}M^0_{static} S^T_{N}
\end{equation}

where the $SL(4,R)$ matrix $S_{N}$ has the following block-diagonal form

\begin{equation}
S_{N} =
\left(%
\begin{array}{cc}
  N & 0 \\
 0 & E_{2\times 2} \\\end{array}%
\right) .
\end{equation}

as one should expect.

Further, we consider the rotating neutral black ring solution

\begin{eqnarray}\label{RBRS}
ds^2_{5} = -{F(y)\over F(x)} \left(dt + C(\nu,\lambda){\cal R}{1+y\over F(y)}d\psi \right)^2 \nonumber\\
+ {{\cal R}^2\over (x-y)^2 }F(x)\left[-{G(y)\over F(y)}d\psi^2 - {dy^2\over G(y)} + {dx^2\over G(x)}
+ {G(x)\over F(x)}d\phi^2 \right]
\end{eqnarray}

where

\begin{equation}
F(x) = 1 + \lambda x ,\,\,\, G(x) = (1-x^2)(1+\nu x),
\end{equation}

and

\begin{equation}
C(\nu,\lambda)= \sqrt{\lambda(\lambda-\nu){1 +\lambda \over 1-\lambda}}.
\end{equation}

The coordinates $x$ and $y$ vary within the ranges

\begin{equation}
 -1\le x \le 1, \,\,\, -\infty < y \le -1,
\end{equation}

and the parameters $\lambda$ and $\nu$ within

\begin{equation}
0<\nu\le \lambda<1.
\end{equation}

In order to avoid conical singularities at $y=-1$ and $x=-1$ the angular variables must be
identified with periodicity

\begin{equation}\label{RRSP}
\Delta \psi = \Delta \phi = 2\pi {\sqrt{1-\lambda}\over 1 -\nu}.
\end{equation}

To avoid  a conical singularity at $x=1$ the parameters $\lambda$ and $\nu$ must be related as

\begin{equation}\label{RRSPC}
\lambda = {2\nu\over 1 +\nu^2}.
\end{equation}

With these choices, the solution has a regular horizon of topology $S^2\times S^1$ at $y=-1/\nu$
and ergosurface of the same topology at $y=-1/\lambda$. Asymptotic spacial infinity is
at $x\to y\to -1$. The static solution is obtained for $\lambda=\nu$ instead of (\ref{RRSPC}).

Following the method described above we make a trivial embedding of the rotating neutral black
ring solution which we shall write in the form

\begin{equation}
ds^2_{6} =f_{44}\left(dt + W_{\psi}d\psi \right)^2
+ {\cal G}_{{\tilde \mu}{\tilde \nu}}dx^{{\tilde \mu}} dx^{{\tilde \nu}} + dx^2_{6}
\end{equation}

where the explicit form of the functions $f_{44}$ and $W_{\psi}$ and the
metric ${\cal G}_{{\tilde \mu}{\tilde \nu}}$ (${\tilde \mu},{\tilde \nu}= 1,2,3,5$) can
be found by comparison with (\ref{RBRS}).

The representative matrix of the embedded solution is given by
\begin{eqnarray}
 M^{0}_{rotating} = \left(%
\begin{array}{cccc}
 f_{44} & 0 & f_{44}W_{\psi} & 0 \\
  0 & 1 & 0 & 0 \\
  f_{44}W_{\psi} & 0 &  f_{\psi\psi} & 0 \\
  0 & 0 & 0 & -f^{-1}
\end{array} %
\right)
\end{eqnarray}

where $f_{\psi\psi}= {\cal G}_{\psi\psi} + f_{44}W^2_{\psi}$ .

New six-dimensional solution is obtained by applying the transformation (\ref{SLT}) with the
matrix $S_{N}$ to $M^{0}_{rotating}$:

\begin{eqnarray}
M_{rotating} = S_{N}M^0_{rotating}S^T_{N} =
\left(%
\begin{array}{cccc}
 {1\over a^2} & -1 & 0 & 0 \\
  -1 & a^2(1 + f_{44}) & af_{44}W_{\psi} & 0 \\
  0 & af_{44}W_{\psi} &  f_{\psi\psi} & 0 \\
  0 & 0 & 0 & -f^{-1}
\end{array} %
\right) .
\end{eqnarray}

Recovering the six-dimensional metric we find

\begin{eqnarray}\label{SDRM}
ds^2_{6} = {f_{44}\over a^2(1 + f_{44})} \left(dt + aW_{\psi}d\psi \right)^2 +
{\cal G}_{{\tilde \mu}{\tilde \nu}}dx^{\tilde \mu} dx^{\tilde \nu} \nonumber \\
+ a^2(1+f_{44}) \left[dx^6 - {dt\over a^2(1 + f_{44})}
+ {f_{44}W_{\psi}\over a(1 + f_{44})}d\psi \right]^2
\end{eqnarray}

After performing Kaluza-Klein reduction we obtain a new rotating EMd solution

\begin{eqnarray}
ds^2_{5} &=& {f_{44}\over \left[a^2(1 + f_{44}) \right]^{2/3}} \left(dt + aW_{\psi}d\psi \right)^2
+
\left[ a^2(1 + f_{44} )\right]^{1/3}{\cal G}_{{\tilde \mu}{\tilde \nu}}
dx^{{\tilde \mu}}dx^{{\tilde\nu}} , \\
A_{\mu}dx^{\mu} &=& - {a^{-2}\over 2(1+ f_{44})}dt + {f_{44}W_{\psi}\over 2a(1 + f_{44})}d\psi, \\
e^{2\varphi} &=& \left[a^2(1+f_{44})\right]^{\sqrt{2/3}}.
\end{eqnarray}

In explicit form the solution reads

\begin{eqnarray}
ds^2_{5} = - \left[a^2\left(1- {F(y)\over F(x)} \right)\right]^{-2/3}{F(y)\over F(x)} \left(dt
+ aC(\nu,\lambda){\cal R}{1+y\over F(y)}d\psi \right)^2 \nonumber \\
+ \left[a^2\left(1- {F(y)\over F(x)} \right)\right]^{1/3} {{\cal R}^2\over (x-y)^2 }F(x)\left[-{G(y)\over F(y)}d\psi^2 - {dy^2\over G(y)} + {dx^2\over G(x)}
+ {G(x)\over F(x)}d\phi^2 \right],\\
A_{\mu}dx^{\mu} = -{1\over 2a^2} { dt \over \left(1- {F(y)\over F(x)} \right)}
- {1\over 2a } {{F(y)\over F(x) }\over \left(1- {F(y)\over F(x)} \right)}
C(\nu,\lambda){\cal R}{1+y\over F(y)}d\psi , \\
e^{2\varphi} = \left[a^2\left(1- {F(y)\over F(x)} \right)\right]^{\sqrt{2/3}}.
\end{eqnarray}

It is worth noting that this EMd solution can be obtained in a simple way via twisted
Kaluza-Klein reduction just as in the case of the four-dimensional, non-asymptotically
flat black holes \cite{CL}. In order to demonstrate this we rewrite the six-dimensional metric
(\ref{SDRM}) in the form

\begin{equation}\label{TTERRS}
ds^2_{6} = f_{44}\left(d{\hat t} + W_{\psi} d\psi\right)^2 +
{\cal G}_{{\tilde \mu}{\tilde \nu}}dx^{{\tilde \mu}} dx^{{\tilde \nu}} + d{\hat x}^2_{6}
\end{equation}

where

\begin{equation}
{\hat t} = a x^{6},\,\,\, {\hat x}^{6} = {t\over a} - ax^6 .
\end{equation}

One immediately recognizes that (\ref{TTERRS}) is the six-dimensional trivial embedding of the
rotating black ring solution. Therefore, the rotating non-asymptotically flat EMd black ring
solution can be obtained from the six dimensional trivial embedding of the
rotating black ring solution  via a twisted Kaluza-Klein reduction with respect to the Killing
vector

\begin{equation}
{\hat \chi} = a \left({\partial \over \partial t} - {\partial \over \partial {x^6}} \right).
\end{equation}

We shall close this section with giving a new five dimensional EMd solution
representing a rotating non-asymptotically flat black hole. It is well known that
the Myers-Perry black hole with rotation in one plane \cite{MP} can be obtained form
the rotating black ring solution as a limiting case \cite{EMP}. Let us consider the same limit
for our EMd solution. Define new parameters

\begin{equation}
m= {2{\cal R}^2\over 1 -\nu}, \,\,\,\, b^2 = 2{\cal R}^2 {\lambda -\nu\over (1-\nu)^2}
\end{equation}

such that they remain finite as $\lambda,\nu \to 1$ and ${\cal R} \to 0$. Also change
the coordinates $(y,x)\to (r,\theta)$

\begin{eqnarray}
x = - 1 + 2\left(1-{b^2\over m}\right){ {\cal R}^2 \sin^2\theta\over r^2 - (m-b^2)\cos^2\theta} ,\\
y = -1 - 2\left(1-{b^2\over m}\right){ {\cal R}^2 \cos^2\theta\over r^2 - (m-b^2)\cos^2\theta} ,
\end{eqnarray}

and rescale the angular coordinates

\begin{equation}
(\psi,\phi) \to \sqrt{{m-b^2\over 2{\cal R}^2}} (\psi,\phi)
\end{equation}

so they now have the canonical periodicity $2\pi$. Then we obtain the following
five-dimensional rotating, nonasymptotically flat, EMd black hole solution

\begin{eqnarray}\label{FDNAFBHS}
ds^2_{5} &=& -\left({\Sigma\over ma^2}\right)^{2/3} \left(1 - {m\over \Sigma} \right)
\left(dt - a{mb\cos^2\theta\over \Sigma -m} d\psi \right)^2 \nonumber \\
&+& \left( {ma^2 \over \Sigma}\right)^{1/3}
\left[\Sigma\left({dr^2\over \Delta}   + d\theta^2\right)
+ {\Delta \cos^2\theta\over 1 -{m\over \Sigma }}d\psi^2 + r^2\sin^2\theta
d\phi^2 \right],\nonumber  \\
A_{\mu}dx^{\mu} &=& -{\Sigma\over 2ma^2 }dt + {b\over 2a}\cos^2\theta d\psi , \\
e^{2\varphi} &=&  \left( {ma^2\over \Sigma}\right)^{\sqrt{2/3}}, \nonumber
\end{eqnarray}

where

\begin{eqnarray}
\Delta = r^2 -m + b^2, \,\,\, \Sigma = r^2 + b^2\sin^2\theta .\nonumber
\end{eqnarray}

\section{Analysis of the solution}

As in the neutral case, in order to avoid singularities at $y=-1$ and $x=-1$ the angular
variable $\psi$ and $\phi$ must be identified with periodicity given by (\ref{RRSP}).
To avoid a conical singularity at $x=1$ the parameters $\lambda$ and $\nu$ must satisfy
the relation (\ref{RRSPC}). Under these conditions the solution has a regular horizon of
topology $S^2\times S^1$ at $y=-1/\nu$ and ergosurface with the same topology at $y=-1/\lambda$.
The regularity of the horizon can be proven by arguments similar to that in \cite{ER1} and \cite{ER2}.
The static solution (\ref{SNFEMDS}) is obtained for $\lambda=\nu$.

The horizon area,  angular  velocity and temperature are found to be

\begin{eqnarray}
{\cal A}_{h} &=& 8\pi^2 a {\cal R}^3 \nu^{3/2}{\sqrt{\lambda(1-\lambda^2)}\over (1-\nu)^2(1+\nu)},\\
\Omega_{h} &=& - {1\over a{\cal R}} \sqrt{\lambda -\nu\over \lambda(1+\lambda)},\\
T &=& {1\over 4\pi a {\cal R}}(1+\nu) \sqrt{{1-\lambda\over \lambda(1+\lambda)\nu }}.
\end{eqnarray}

The surface $y=-1/\nu$ is a Killing horizon of the Killing vector

\begin{equation}
{\cal K} = {\partial\over \partial t }
+ {\sqrt{1-\lambda}\over 1-\nu } \Omega_{h} {\partial\over \partial \psi} =
{\partial\over \partial t }
+  \Omega_{h} {\partial\over \partial {\tilde \psi}}
\end{equation}

where ${\tilde \psi}= {1-\nu\over \sqrt{1-\lambda}}\psi$ (see below). The temperature
is

\begin{equation}
T = {\kappa \over 2\pi}
\end{equation}

and the surface gravity $\kappa$ associated with the Killing vector ${\cal K}$ is defined as usual.

The spacial infinity is reached as $x\to y\to -1$. The asymptotic behaviour of the solution
is

\begin{eqnarray}\label{RBRAM}
ds^2_{5}&\approx & - \left[a^{-2} {1-\nu\over 2\lambda}{r^2\over {\cal R}^2} \right]^{2/3}
dt^2   \\
&+& \left[a^{-2} {1-\nu\over 2\lambda}{r^2\over {\cal R}^2} \right]^{-1/3}
\left[dr^2  + r^2d\theta^2 +
r^2\cos^2\theta d{\tilde \psi}^2 +   r^2\sin^2\theta d{\tilde \phi}^2\right]  \nonumber,
\end{eqnarray}

\begin{eqnarray}\label{RBRAGP}
A_{\mu}dx^{\mu} &\approx &
- {1\over 2}\left[a^{-2} {1-\nu\over 2\lambda}{r^2\over {\cal R}^2}\right]dt +
{1\over 2a} {C(\nu,\lambda)\over \lambda}{\sqrt{1-\lambda}\over 1-\nu }{\cal R}\cos^2\theta
d{\tilde \psi} ,
\end{eqnarray}

\begin{eqnarray}\label{RBRAD}
e^{2\varphi} &\approx &
\left[a^{-2}{1-\nu\over 2\lambda}{r^2\over {\cal R}^2}\right]^{-\sqrt{2/3}} ,
\end{eqnarray}

where the new coordinates $r$, $\theta$, ${\tilde \psi}$ and ${\tilde \phi}$  are given by

\begin{eqnarray}
r\cos\theta &=& \sqrt{{1-\lambda\over 1 -\nu}} {\cal R} {\sqrt{y^2-1}\over x-y}, \\
r\sin\theta &=& \sqrt{{1-\lambda\over 1 -\nu}}{\cal R} {\sqrt{1-x^2}\over x-y}, \\
{\tilde \psi} &=& {1-\nu\over \sqrt{1-\lambda}}\psi, \\
{\tilde \phi} &=& {1-\nu\over \sqrt{1-\lambda}}\phi .
\end{eqnarray}

The angular coordinates ${\tilde \psi}$ and ${\tilde \phi}$  are with canonical periodicity
$2\pi$. The asymptotic form of the solution shows obviously that the solutions is not
asymptotically flat. This asymptotic matches the asymptotic of the Chan-Horne-Mann solution \cite{CHM} for
$\alpha^2=8/3$ and the asymptotic of the black hole solution (\ref{FDNAFBHS}). The curvature scalars  vanish
as $r\to \infty$ (see Appendix \ref{RKSSS})

\begin{equation}
R\sim r^{-4/3}, \,\,\,\, R_{\mu\nu\alpha\beta}R^{\mu\nu\alpha\beta} \sim r^{-8/3}
\end{equation}

and the spacetime is well-behaved there.

In order to calculate the mass and angular momentum of the solution
we use the  quasilocal formalism (see Appendix \ref{QLF}). After a
long algebra we find

\begin{eqnarray}
M &=& {\pi{\cal R}^2\over 2} {\lambda \over 1-\nu} ,\\
J_{\psi} &=& - a {\pi {\cal R}^3\over 2} {\sqrt{\lambda(\lambda-\nu)(1+\lambda)}\over (1-\nu)^2}.
\end{eqnarray}

The electric charge is defined by

\begin{equation}
Q = {1\over 8\pi} \oint_{\infty}e^{2\alpha\varphi}F_{\mu\nu}d\Sigma^{\mu\nu}
\end{equation}

and we find

\begin{equation}
Q = - \pi a^2  {\cal R}^2 {\lambda \over 1-\nu}.
\end{equation}

The electric potential in the horizon rest frame evaluated at the horizon is

\begin{equation}
{\Phi}_{h} = {\cal K}^{\mu}A_{\mu}.
\end{equation}

This potential can be chosen by adding an appropriate constant so that
the following Smarr-type relation to be satisfied

\begin{equation}
M = {3\over 2}\left( T{{\cal A}_{h}\over 4 } + \Omega_{h}J_{\psi} \right) + \Phi_{h}Q .
\end{equation}

Moreover, with the same choice of the potential $\Phi_{h}$ one can show that the first law

\begin{equation}
dM = T d\left({{\cal A}_{h}\over 4 } \right) + \Omega_{h}dJ_{\psi} + \Phi_{h}dQ
\end{equation}

is satisfied\footnote{Note that the parameter $a$ is kept fixed}.

The physical parameters and the thermodynamics relations of the black hole solution
(\ref{FDNAFBHS}) are obtained as a limiting case of those for the black rings.

\section{Conclusion}

In this paper we derived new rotating, non-asymptotically flat black ring solutions
in a five-dimensional EMd theory which is  dimensional reduction of a six-dimensional
Kaluza-Klein theory. As a limiting case, new five-dimensional, non-asymptotically flat
black holes with rotation in one plane were obtained, too.
The properties of the solutions were computed. A Smarr-like relation
was found. It was shown that the fitst law of black hole thermodynamics is satisfied.
Here we considered black rings rotating in $\psi$-direction.
The construction of black rings with rotation in the azimuthal direction of the $S^2$
is straightforward.

Finally, it would be interesting to find rotating, asymptotically and non-asymptotically flat
black rings for arbitrary dilaton coupling parameter $\alpha$. This task, however,
requires a systematic study of the stationary EMd equations in five (and more) dimensions. We will
address this problem in a future work.

\section*{Acknowledgements}

This work was partially supported by the Bulgarian National Science Fund under Grant MUF04/05 (MU 408).
The author would like
to thank the Bogoliubov Laboratory of Theoretical Physics (JINR) for their kind hospitality.

\appendix

\section{Ricci and Kretschmann scalars}\label{RKSSS}

In this Appendix we give the asymptotic behaviour of the Ricci and Kretschmann scalars for the dilaton black
ring spacetimes.

\subsection{Ricci and Kretschmann scalars for the static black rings}

The Ricci scalar is given by

\begin{eqnarray}
R = {3\over 2(1+ \alpha^2_{5}) } \left[{\alpha^2_{5}\over 1+ \alpha^2_{5} } {F(y)\over F(x) } -{1\over 2}  \right]
{\left[F(y)(1-x^2) + F(x)(y^2-1) \right]\over {\cal R}^2 F(x)} Z^{- {1\over 1 + \alpha^2_{5} }}
\end{eqnarray}

where

\begin{equation}
Z = a^2\left[1 - {F(y)\over F(x)} \right].
\end{equation}

The Ricci scalar is regular on the horizon and everywhere outside the horizon. For $r\to \infty$ we find

\begin{equation}
R\simeq {3(\alpha^2_{5} - 1)\over 2(1 + \alpha^2_{5})^2 } {(1-\lambda) \over \lambda a^2 {\cal R}^2 }
\left[a^2 {2\lambda\over 1-\lambda } {{\cal R}^2 \over r^2} \right] ^{\alpha^2_{5}\over 1+ \alpha^2_{5} } \sim
r^{-{2\alpha^2_{5}\over 1+ \alpha^2_{5}}}
\end{equation}

when $\alpha^2_{5} \ne 1$ and

\begin{equation}
R \sim r^{-3}
\end{equation}

when $\alpha^2_{5}=1$. This asymptotic behaviour can be also  obtained directly form the asymptotic form of the metric (\ref{SBRAM}).

The Kretschmann scalar is difficult to calculate even by computer. However, its asymptotic behaviour can be found from the asymptotic form of the metric (\ref{SBRAM}). Making use of the formulas given in \cite{DR} we obtain

\begin{equation}
R_{\mu\nu\alpha\beta}R^{\mu\nu\alpha\beta} \sim r^{-{4\alpha^2_{5}\over 1+ \alpha^2_{5}}}.
\end{equation}

\subsection{Ricci and Kretschmann scalars for the rotating black rings}

The Ricci scalar is given by the expression

\begin{eqnarray}
R = {1\over 3}{ Z^{-1/3}\over {\cal R}^2 F^3(x) } \left[G(x)F^2(y) - G(y)F^2(x) \right]  \nonumber \\+
{Z^{-1/3}\over 4 {\cal R}^2 F^3(x) } \left[ -G(x)F(y)F(x) + 3G_{E}\left[F(x)-F(y)\right]F(x) + G_{E}F^2(y) \nonumber \right.\\ \left. + {1\over \lambda }\left[\nu(1-y^2) + ({1\over \lambda } -y)(1 -{\nu\over \lambda })
 \right] F^3(x) \right] \nonumber
\end{eqnarray}

where

\begin{equation}
G_{E}= G(y)|_{y=-{1\over \lambda}}= (1- {1\over \lambda^2}) (1-{\nu\over \lambda }).
\end{equation}

The Ricci scalar curvature is well-behaved on and outside the horizon. The asymptotic behavior of the Ricci scalar for $r\to \infty$ is

\begin{equation}
R \simeq {1\over 6 } {(1-\nu) \over \lambda a^2 {\cal R}^2} \left[a^2 {2\lambda\over 1-\nu } {{\cal R}^2\over r^2}\right]^{2/3} \sim  r^{-4/3}.
\end{equation}

Using the asymptotic form of the metric (\ref{RBRAM}) we find

\begin{equation}
R_{\mu\nu\alpha\beta}R^{\mu\nu\alpha\beta} \sim r^{-8/3}.
\end{equation}

\section{Quasilocal formalism}\label{QLF}

Here we briefly discuss the quasilocal formalism in EMd gravity \cite{CGL}.
The spacetime metric can be decomposed into the form

\begin{equation}
ds^2 = - N^2dt^2 + \chi_{ij}(dx^i + N^{i}dt)(dx^j + N^{j}dt)
\end{equation}

where $N$ is the lapse function and $N^{i}$ is the shift vector.

This decomposition means that the spacetime is foliated by spacelike surfaces $\Sigma_{t}$ of metric
$\chi_{\mu\nu} = g_{\mu\nu} + u_{\mu}u_{\nu}$, labeled by a time coordinate $t$ with a unit normal vector
$u^{\mu} = - N\delta^{\mu}_{0}$. A timelike vector $\upsilon^{\mu}$, satisfying $\upsilon^{\mu}\nabla_{\mu}t=1$,
is decomposed into the lapse function and shift vector as $\upsilon^{\mu}= Nu^{\mu} + N^{\mu}$.
The spacetime boundary consists in the initial surface $\Sigma_{i}$ ($t=t_{i}$) and the final surface $\Sigma_{f}$
($t=t_{f}$) and a timelike surface ${\cal B}$ to which the vector $u^{\mu}$ is tangent. The surface ${\cal B}$
is foliated by $(D-2)$-dimensional surfaces $S^{r}_{t}$, of metric $\sigma_{\mu\nu}= \chi_{\mu\nu} - n_{\mu}n_{\nu}$,
which are intersection of $\Sigma_{t}$ and ${\cal B}$. The unit spacelike outward normal to $S^{r}_{t}$, $n_{\mu}$,
is orthogonal to $u^{\mu}$.

The quasilocal energy and angular momentum are given by

\begin{eqnarray}
E = {1\over 8\pi} \int_{S^{r}_{t}} \sqrt{\sigma} \left[N(k-k_{0}) + {n_{\mu}p^{\mu\nu}N_{\nu}\over \sqrt{\chi}} \right] d^{D-2}x \nonumber \\ +
{1\over 4\pi}\int_{S^{r}_{t}} A_{0} \left({\hat \Pi}^{j} - {\hat \Pi}_{0}^{j} \right)n_{j}d^{D-2}x ,\\
J_{i} = - {1\over 8\pi} \int_{S^{r}_{t}} {n_{\mu}p^{\mu}_{i}\over \sqrt{\chi} } \sqrt{\sigma}d^{D-2}x
- {1\over 4\pi} \int_{S^{r}_{t}} A_{i}{\hat \Pi}^{j}n_{j}d^{D-2}x .
\end{eqnarray}

Here $k= - \sigma^{\mu\nu}D_{\nu}n_{\mu}$ is the trace of the extrinsic curvature of $S^{r}_{t}$ embedded in $\Sigma_{t}$
and $D_{\nu}$ is the coderivative with respect to the metric $\chi_{\mu\nu}$. The momentum variable $p^{ij}$
conjugated to $\chi_{ij}$ is given by

\begin{equation}
p^{ij} = \sqrt{\chi}\left(\chi^{ij}K - K^{ij} \right)
\end{equation}

where

\begin{equation}
K_{ij} = - {1\over 2N}\left( {\partial \chi_{ij}\over \partial t} - 2D_{(i}N_{j)} \right)
\end{equation}

is the extrinsic curvature of $\Sigma_{t}$ and $K$ is its trace.
The quantity ${\hat \Pi}^{j}$ is defined by

\begin{equation}
{\hat \Pi}^{j} = - {\sqrt{\sigma}\over \sqrt{\chi}} \sqrt{-g}e^{2\alpha\varphi}F^{0j}.
\end{equation}

The quantities with the subscript "0" are those associated with the background. Detail discussion of the quasilocal
formalism can be found in \cite{CGL}.

\subsection{The mass of the static black rings}

The mass of the solution is defined as the limit $M= \lim_{r\to \infty} E(r)$. Keeping only the leading asymptotic
monopole and dipole contributions for $r\to \infty$ we have

\begin{eqnarray}
\sigma_{ab}dx^a dx^b \simeq  Z^{1\over 1 + \alpha^2_{5}} r^2 d\Omega^2_{3} ,\\
N \simeq Z^{-{1\over 1 + \alpha^2_{5}}} \left(1 - {\lambda\over 1-\lambda }{{\cal R}^2\over r^2} \right),\\
n^{r} \simeq Z^{-{1\over 2(1 + \alpha^2_{5})}} \left(1 -{\lambda\over 1-\lambda }{{\cal R}^2\over r^2} \right),
\end{eqnarray}

where

\begin{equation}
Z \simeq a^2 {2\lambda\over 1-\lambda} {{\cal R}^2\over r^2}
\end{equation}

for $r\to \infty$. For the asymptotic extrinsic curvature we obtain

\begin{equation}
k \simeq  -  {3\alpha^2_{5} \over 1 + \alpha^2_{5} } {Z^{-{1\over 2(1+ \alpha^2_{5})} }\over r }\left(1 -{\lambda\over 1-\lambda }{{\cal R}^2\over r^2} \right).
\end{equation}

The natural background solution is given by (\ref{SBRAM})-(\ref{SBRAP}) and the corresponding extrinsic curvature is

\begin{equation}
k_{0} \simeq  - {3\alpha^2_{5} \over 1 + \alpha^2_{5} }{Z^{-{1\over 2(1+ \alpha^2_{5})} }\over r }.
\end{equation}

Then we find

\begin{eqnarray}
M &=& {1\over 8\pi }{3\alpha^2_{5}\over 1+ \alpha^2_{5} } \lim_{r\to \infty} r^2\left[ {\lambda\over 1-\lambda }{{\cal R}^2\over r^2 } \right] \int^{\theta=\pi/2}_{\theta=0} \int^{{\tilde \psi}=2\pi}_{{\tilde \psi}=0} \int^{{\tilde \phi}=2\pi\sqrt{{1-\lambda\over 1\pm \lambda }}}_{{\tilde \phi}=0} \cos\theta\sin\theta d\theta d{\tilde\psi }
d{\tilde\phi } \nonumber \\
&=& {\alpha^2_{5}\over 1+ \alpha^2_{5} } {3\pi{\cal R}^2\over 4 } {\lambda \over \sqrt{(1-\lambda)(1\pm \lambda)} }.
\end{eqnarray}

\subsection{The mass and the angular momentum of the rotating black rings}

Keeping only the leading asymptotic
monopole and dipole contributions for $r\to \infty$ we have

\begin{eqnarray}
\sigma_{ab}dx^a dx^b \simeq  Z^{1\over 3} r^2 d\Omega^2_{3} ,\\
N \simeq Z^{-{1\over 3}} \left(1 - {\lambda\over 1-\nu }{{\cal R}^2\over r^2} \right),\\
n^{r} \simeq Z^{-{1\over 6}} \left(1 -{\lambda\over 1-\nu }{{\cal R}^2\over r^2} \right), \\
N^{{\tilde \psi}} \simeq -2 a C(\nu,\lambda) {\sqrt{1-\lambda}\over (1-\nu)^2 } {Z^{-1}{\cal R}^2 \over r^4},
\end{eqnarray}

where

\begin{equation}
Z \simeq a^2 {2\lambda\over 1-\nu} {{\cal R}^2\over r^2} .
\end{equation}

The natural background solution is given by (\ref{RBRAM})-(\ref{RBRAD}). In this case the integral

\begin{equation}
{1\over 4\pi}\int_{S^{r}_{t}} A_{0} \left({\hat \Pi}^{j} - {\hat \Pi}_{0}^{j} \right)n_{j}d^{3}x
\end{equation}

is zero. It can be shown that the  integral

\begin{equation}
{1\over 8\pi} \int_{S^{r}_{t}} \sqrt{\sigma} \left[ {n_{\mu}p^{\mu\nu}N_{\nu}\over \sqrt{\chi}} \right] d^{3}x
\sim {1\over r^2}
\end{equation}

and therefore it does not contribute  to the mass. Consequently, only the integral

\begin{equation}
{1\over 8\pi} \int_{S^{r}_{t}} \sqrt{\sigma} N(k-k_{0}) d^{3}x
\end{equation}

contributes  to the mass of the solution. The asymptotic extrinsic curvatures are

\begin{equation}
k \simeq -2 {Z^{-{1\over 6}}\over r } \left(1 - {\lambda\over 1-\nu} {{\cal R}^2\over r^2}\right),
\end{equation}

\begin{equation}
k_{0} \simeq -2 {Z^{-{1\over 6}}\over r } .
\end{equation}

Substituting into the integral we find

\begin{eqnarray}
M &=& {1\over 4\pi} \lim_{r\to \infty} r^2\left[ {\lambda\over 1-\nu }{{\cal R}^2\over r^2 } \right] \int^{\theta=\pi/2}_{\theta=0} \int^{{\tilde \psi}=2\pi}_{{\tilde \psi}=0} \int^{{\tilde \phi}=2\pi}_{{\tilde \phi}=0} \cos\theta\sin\theta d\theta d{\tilde\psi }
d{\tilde\phi } \nonumber \\
&=& {\pi{\cal R}^2\over 2 } {\lambda \over 1-\nu }.
\end{eqnarray}

The angular momentum is

\begin{eqnarray}
J_{\psi} = - {1\over 8\pi} \lim_{r\to\infty }\int_{S^{r}_{t}} {n_{\mu}p^{\mu}_{{\tilde \psi}}\over \sqrt{\chi} } \sqrt{\sigma}d^{3}x
- {1\over 4\pi} \lim_{r\to\infty }\int_{S^{r}_{t}} A_{{\tilde \psi}}{\hat \Pi}^{j}n_{j}d^{3}x .
\end{eqnarray}

The first integral reduces to

\begin{equation}
- {1\over 8\pi} \lim_{r\to\infty }\int_{S^{r}_{t}} {n_{\mu}p^{\mu}_{{\tilde \psi}}\over \sqrt{\chi} } \sqrt{\sigma}d^{3}x = {1\over 8\pi} \lim_{r\to\infty } \int_{S^{r}_{t}} n_{r}K^{r}_{{\tilde \psi}}\sqrt{\sigma}d^3x.
\end{equation}

For $r\to \infty$ we obtain

\begin{equation}
n_{r}K^{r}_{{\tilde \psi}}\sqrt{\sigma} \simeq -2a C(\nu,\lambda) {\sqrt{1-\lambda}\over (1-\nu)^2}{\cal R}^3
\cos^3\theta \sin\theta.
\end{equation}

Therefore we find

\begin{eqnarray}
&{1\over 8\pi}& \lim_{r\to\infty } \int_{S^{r}_{t}} n_{r}K^{r}_{{\tilde \psi}}\sqrt{\sigma}d^3x = \nonumber \\
 &-& {1\over 4\pi} a C(\nu,\lambda) {\sqrt{1-\lambda}\over (1-\nu)^2}{\cal R}^3
\int^{\theta=\pi/2}_{\theta=0} \int^{{\tilde \psi}=2\pi}_{{\tilde \psi}=0} \int^{{\tilde \phi}=2\pi}_{{\tilde \phi}=0} \cos^3\theta\sin\theta d\theta d{\tilde\psi }d{\tilde \phi} \nonumber \\
&=& - {\pi a\over 4 } C(\nu,\lambda) {\sqrt{1-\lambda}\over (1-\nu)^2}{\cal R}^3 .
\end{eqnarray}

In order to compute the second integral in $J_{\psi}$ we first find $A_{{\tilde \psi}}$ and ${\hat \Pi}^{r}n_{r}$
for $r\to \infty$:

\begin{eqnarray}
A_{{\tilde \psi}} \simeq {1\over 2a} {C(\nu,\lambda)\over \lambda } {\sqrt{1-\lambda}\over 1-\nu } {\cal R} \cos^2\theta ,\\
{\hat \Pi}^{r}n_{r} \simeq a^2 {2\lambda\over 1-\nu} {\cal R}^2 \cos\theta\sin\theta .
\end{eqnarray}

Then we find

\begin{eqnarray}
- &{1\over 4\pi}& \lim_{r\to\infty }\int_{S^{r}_{t}} A_{{\tilde \psi}}{\hat \Pi}^{j}n_{j}d^{3}x
 \nonumber \\
&=& - {1\over 4\pi} a C(\nu,\lambda) {\sqrt{1-\lambda}\over (1-\nu)^2 } {\cal R}^3
\int^{\theta=\pi/2}_{\theta=0} \int^{{\tilde \psi}=2\pi}_{{\tilde \psi}=0} \int^{{\tilde \phi}=2\pi}_{{\tilde \phi}=0} \cos^3\theta\sin\theta d\theta d{\tilde\psi }d{\tilde \phi} \nonumber \\
&=& - {\pi a\over 4 } C(\nu,\lambda) {\sqrt{1-\lambda}\over (1-\nu)^2}{\cal R}^3 .
\end{eqnarray}

Finally,  for the angular momentum  we obtain

\begin{eqnarray}
J_{\psi} = - {\pi a\over 2 } C(\nu,\lambda) {\sqrt{1-\lambda}\over (1-\nu)^2}{\cal R}^3 =
- a{\pi {\cal R}^3 \over 2} {\sqrt{\lambda(\lambda-\nu)(1+\lambda)}\over (1-\nu)^2 } .
\end{eqnarray}


\begin{thebibliography}{tbds}

\bibitem{ER1} R.~Emparan~and~H.~Reall, ~Phys.~Rev.~{\bf D65},~084025~(2002).
\bibitem{ER2} R.~Emparan~and~H.~Reall, ~Phys.~Rev.~Lett.~{\bf 88},~101101~(2002).
\bibitem{E} H.~Elvang,~Phys.~Rev.~{\bf D68},~124016~(2003).
\bibitem{EEMR1} H.~Elvang,~R.~Emparan,~D.~Mateos~and~H.~Reall~Phys.~Rev.~Lett.~{\bf 93},~211302~(2004).
\bibitem{GG1} J.~Gauntlett~and~J.~Gutowski, ~Phys.~Rev.~{\bf D71},~025013~(2005).
\bibitem{GG2} J.~Gauntlett~and~J.~Gutowski, ~Phys.~Rev.~{\bf D71},~045002~(2005).
\bibitem{EE} H.~Elvang,~ R.~Emparan,~JHEP~{\bf 0311},~035~(2003).
\bibitem{EEMR2} H.~Elvang,~R.~Emparan,~D.~Mateos~and~H.~Reall~Phys.~Rev.~{\bf D71},~024033~(2004).
\bibitem{BWW} I.~Bena,~C.~Wang~and~N.~ Warner, hep-th/0411072.
\bibitem{HR}  G.~Horowitz~and~H. Reall,~Class.~Quant.~Grav.~{\bf 22},~1289~(2005)
\bibitem{BWW} I.~Bena,~P.~Kraus~and~N.~ Warner, hep-th/0504142.
\bibitem{O}   M.~Ortaggio,~JHEP~{\bf 0505},~048~(2005).
\bibitem{MI}  T.~Mishima~and~H.~Iguchi,~hep-th/0504018.
\bibitem{P}   P.~Figueras,~hep-th/0505244.
\bibitem{IU} D.~Ida~and~Y.~Uchida, ~Phys.~Rev.~{\bf D68},~104014~(2003).
\bibitem{EMP} R.~Emparan,~JHEP~{\bf 0403},~064~(2004).
\bibitem{AR}  D.~Astefanesei~and~E.~Radu,~hep-th/0509144 .
\bibitem{KL} H.~Kunduri~and~J.~Lucietti, ~Phys.~Lett.~{\bf B609},~143~(2005).
\bibitem {Y} S.~Yazadjiev,~hep-th/0507097
\bibitem{PW} S.~Poletti~and~D.~Wiltshire,~Phys.~Rev.~{\bf D50},~7260~(1994);~{\bf D52},~3753(E)~(1995).
\bibitem{CHM} K.~Chan,~J.~Horne~and~R.~Mann,~Nucl.~Phys.~{\bf B447},~441~(1995).
\bibitem{CJS} R.~Cai,~J.~Ji~and~K.~Soh,~Phys.~Rev.~{\bf D57},~6547~(1998).
\bibitem{CZ} R.~Cai~and~Y.~Zhang,~Phys.~Rev.~{\bf D54},~4891~(1996).
\bibitem{CGL} G.~Clement,~D.~Gal'tsov~and~C.~Leygnac,~Phys.~Rev.~{\bf D67},~024012~(2003).
\bibitem{CL} G.~Clement~and~C.~Leygnac,~Phys.~Rev.~{\bf D70},~084018~(2004).
\bibitem{CW} R.~Cai~and~A.~Wang,~Phys.~Rev.~{\bf D70},~084042~(2004).
\bibitem{Y1} S.~Yazadjiev,~Class.~Quant.~Grav.~{\bf 22},~3875~(2005).
\bibitem{GR} D.~Gal'tsov~and~O.~Rytchkov,~Phys.~Rev.~{\bf D58},~122001~(1998).
\bibitem{M}  D.~Maison,~Gen.~Rel.~Grav.~{\bf 10},~717~(1979).
\bibitem{DM} P.~Dobiasch~and~D.~Maison,~Gen.~Rel.~Grav.~{\bf 14},~231~(1982).
\bibitem{GW} G.~Gibbons~and~D.~Wiltshire,~Ann.~Phys.~{\bf 167},~201~(1985).
\bibitem{CLEM}G.~Clement,~Gen.~Rel.~Grav.~{\bf 18},~137~(1986).
\bibitem{FZB}V.~Frolov,~A.~Zelnikov~and~U.~Bleyer,~Ann.~Phys.~{\bf 44},~371~(1987).
\bibitem{HH} J.~Horne~and~G.~Horowitz,~Phys.~Rev.~{\bf D46},~1340~(1992).
\bibitem{MP} R.~Myers~and~M.~Perry,~Ann.~Phys.~{\bf 172},~304~(1986).
\bibitem{DR} S.~Deser~and~A.~Ryzhov,~Class.~Quant.~Grav.~{\bf 22},~3315~(2005).


\end{thebibliography}
\end{document}